\documentclass[%
reprint,
preprintnumbers,
amsmath,amssymb,
]{revtex4-1}

\usepackage{graphicx}% Include figure files
\usepackage{dcolumn}% Align table columns on decimal point
\usepackage{bm}% bold math
\usepackage{amsmath,amssymb,amsthm}
\usepackage{graphicx}
\usepackage{amsfonts,graphics,epsfig, epstopdf}
\usepackage{color}
\usepackage{enumerate}
\usepackage{fixltx2e}
\begin{document}

%\preprint{APS/123-QED}

\title{\textbf{Constrained-search density functional study of quantum transport in two-dimensional vertical heterostructures}}% Force line breaks with \\

% 자동 세팅된 것은, 표식이 asterisk, dagger 순서
\author{Han Seul Kim}

 \altaffiliation[Current address: ]{Division of National Supercomputing Research and Development, Korea Institute of Science and Technology Information (KISTI), 245 Daehak-ro, Yuseong-gu, Daejeon 34141, Republic of Korea}

 \affiliation{School of Electrical Engineering and Graduate School of Energy, Environment, Water, and Sustainability, Korea Advanced Institute of Science and Technology (KAIST), 291 Daehak-ro, Yuseong-gu, Daejeon 34141, Republic of Korea}

\author{Yong-Hoon Kim}%
 \email{y.h.kim@kaist.ac.kr}
\affiliation{School of Electrical Engineering and Graduate School of Energy, Environment, Water, and Sustainability, Korea Advanced Institute of Science and Technology (KAIST), 291 Daehak-ro, Yuseong-gu, Daejeon 34141, Republic of Korea}%


\begin{abstract}
Based on a microcanonical picture that maps the steady-state quantum transport process to a drain-to-source excitation, we develop a constrained-search density functional formalism for finite-bias quantum transport calculations. 
By variationally minimizing the total energy of an electrode-channel-electrode system without introducing separate bulk electrode information, ambiguities in identifying its nonequilibrium electronic structure under a bias is reduced and finite electrode cases can be naturally treated. 
We apply the approach to vertically stacked van der Waals heterostructures made of a hexagonal boron nitride (hBN) channel sandwiched by single-layer graphene electrodes, which so far could not be treated within first-principles calculations. 
We find that the experimentally observed negative differential resistance originates from the hBN defect-mediated hybridizations between two graphene states, and concurrently obtain a high-bias linear current increase that was not captured in previous semiclassical treatments.
Going beyond the capability of existing \textit{ab initio} nonequilibrium quantum transport simulation methods, the developed formalism will provide valuable atomistic information in the development of next-generation nanodevices.

\begin{description}
%\item[DOI] DOI
\item[Subject Areas] 
%Computational Physics, \\
%Condensed Matter Physics - Graphene, \\
%Materials Science, \\
%Nanophysics, \\
%Quantum Physics % 지정된 항목들 중에서 골라 적어야 함.
Quantum transport $|$ Constrained-search density functional theory $|$ Graphene $|$ Vertical transistor $|$ Negative differential resistance
\end{description}
\end{abstract}
\maketitle
%\tableofcontents

% PRX 논문들을 보면, Section은 논문마다 달기도 하고 달지 않기도 함.
\section{\label{sec:level1}Introduction}
%% Introduction
Density functional theory (DFT) in the standard form cannot be applied to nonequilibrium quantum electron transport phenomena, thus in the last decade or so the DFT-based nonequilibrium Green's function (NEGF) formalism has been established as the standard approach for first-principles quantum transport calculations \cite{Datta2005,DiVentra2008}. 
While successful, the DFT-NEGF approach suffers from several shortcomings due the the Landauer picture invoked in its practical realization. 
The device is viewed within the Landauer picture as an open system with an external battery as the source of the current flow, and to implement this viewpoint one replaces the Hamiltonian and related matrix elements corresponding to the \emph{semi-infinite} electrode regions with those from separate \emph{infinite} bulk calculations.

So, DFT-NEGF requires the electrodes to be repeated semi-infinitely, making junction models based on finite-dimension electrodes intrinsically non-tractable. This unfortunately implies that the recently developed transistors constructed by vertically stacking two-dimensional (2D) semiconductor materials and single-layer graphene electrodes \cite{Britnell2012_SC,Britnell2013,Jena2013,Li2016,Iannaccone2018} cannot be treated within DFT-NEGF.

Due to the fundamental limitation of the DFT-NEGF method, simulations of transistors based on 2D heterostructures are currently performed by replacing single-layer graphene (\textcolor[rgb]{0,0,1}{Fig. 1\textit{A}}) with infinite-layer graphite electrodes (\textcolor[rgb]{0,0,1}{Fig. 1\textit{B}}) or resorting to semi-classical approaches such as the Bardeen transfer Hamiltonian formalism. Both directions are clearly undesirable in that graphene and graphite electrodes behave differently in the former case \cite{Mak2010, Kang2016} and the possibility of interpreting and predicting effects that involve atomistic details in an \textit{ab initio} manner will be eliminated in the latter \cite{Bardeen1961,Tersoff1985,Feenstra2012,Zhao2013,Campbell2015}. 

In this work, we report the development of a multi-space constrained-search formulation of DFT for nonequilibrium electronic structure calculations and its applications to graphene-based vertical tunneling transistors.
The multi-space constrained-search DFT (MS-DFT) formalism is established by adopting the micro-canonical picture, in which one considers finite-sized electrodes in contrast to semi-infinite electrodes within the grand canonical picture of NEGF.
A key conceptual step we newly introduce is to view the finite-bias \textit{quantum transport process} as the \textit{drain-to-source electronic excitation}. 
The resulting MS-DFT provides an alternative framework to the standard DFT-NEGF scheme for first-principles nonequilibrium quantum transport calculations and can be straightforwardly implemented within an existing DFT code. 
Applying MS-DFT to vertical 2D van der Waals (vdW) heterostructures made of hexagonal boron nitrides (hBNs) sandwiched by graphene monolayers, we obtain for the first time the negative differential resistance (NDR) and subsequent increasing linear current in a first-principles manner and clarify their atomistic origins. 

\section{\label{sec:2}Formulation and implementation of MS-DFT}
\subsection{\label{sec:2-1} MS-DFT for steady-state quantum transport}

To establish MS-DFT, we first switch from the standard grand-canonical or Landauer picture to the micro-canonical one, in which electrical currents can be viewed as the long-lived discharging of large but finite capacitors. 
Such an approach was initially explored by Di Ventra and Todorov \cite{DiVentra2004}, but they focused on combining it with time-dependent DFT to study transient (rather than steady-state) electron dynamics \cite{Bushong2005}.
Next, we divide the junction into left electrode (\textit{L}), channel (\textit{C}), and right electrode (\textit{R}) regions, and trace the spatial origins of a wave function $\Psi$ to \textit{L}, \textit{C}, or \textit{R}. 
At the zero-bias limit, together with one global Fermi level, they collectively give the ground-state density  $\rho_{0}(\vec{r})=\rho_{0}^{L}(\vec{r})+\rho_{0}^{C}(\vec{r})+\rho_{0}^{R}(\vec{r})$.
It is assumed that we have a semiconducting (or insulating) region within \textit{C}, while the \textit{L} and \textit{R} regions are metallic. 
Finally, we view the finite applied bias voltage $V_b = (\mu_R - \mu_L)/e$, where $\mu_R$ ($\mu_L$) indicates the chemical potential of the region \textit{R} (\textit{L}), 
as the excitation from the states spatially belonging to the drain electrode \textit{L} to those of the source electrode \textit{R}, and apply a constrained search to those spatially ``excited" states with density $\rho_{k}$. 
Namely, we establish the mapping of the transport problem to the optical one, and in doing so generalize the variational (time-independent) excited-state DFT, which is formally well-established by Levy-Nagy \cite{Levy1999,Ayers2009} and G\"orling \cite{Gorling1999}, to the multi-space (from the drain to source electrode) excitation case. 
In other words, the role of light in the time-independent DFT is played by the external battery in the MS-DFT, and it is mathematically embodied by a multi-space constraint. 

Then, given the ground state with the total energy $E_{0}$ and density $\rho_{0}$, the governing equation of the MS-DFT becomes a constrained search of the total energy minimum of the excited state \textit{k} with the density $\rho_{k}(\vec{r})=\rho_{k}^{L}(\vec{r})+\rho_{k}^{C}(\vec{r})+\rho_{k}^{R}(\vec{r})$,
%
%\begin{equation}
%\begin{multilined}
\begin{multline}
	E_{k} = 
	%YHK: min_{\rho}
	\underset{\rho}{\textrm{min}}\left \{ \int v(\vec{r}) \rho(\vec{r}) d^{3} \vec{r} + F\left [ \rho_{k}^{L}, \rho_{k}^{C}, \rho_{k}^{R}, \rho_{0} \right ] \right \} \\
	=  \int v(\vec{r}) \rho(\vec{r}) d^{3} \vec{r} + F\left [ \rho_{k}, \rho_{0} \right ],
	\label{eq1}
\end{multline}
%\end{multilined}
%\end{equation}
%
with the universal functional
%
%\begin{equation}
\begin{multline}
	F\left [ \rho_{k}, \rho_{0} \right ] = \\
	%YHK: min_{\Psi^{L/C/R}\rightarrow \rho_{k} }
	\underset{\Psi^{L/C/R}\rightarrow \rho_{k}}{\textrm{min}} 
	\left \langle \Psi^{L/C/R}\left | \widehat{T}+\widehat{V}_{ee} \right | \Psi^{L/C/R} \right \rangle,
	\label{eq2}
\end{multline}
%\end{equation}
%
where the spatially-resolved $\Psi^{L/C/R}$ are understood to be restricted to the states that satisfy the bias constraint $eV_b = \mu_{R} - \mu_{L}$ and are orthognal to the first $k-1$ excited states. By solving the corresponding KS equations \cite{Levy1999,Gorling1999,Ayers2009},
\begin{equation}
	\left [ \widehat{h}^{0}_{KS} + \Delta v_{Hxc} (\vec{r}) \right ] \psi_{i} (\vec{r}) = \epsilon_{i} \psi_{i}(\vec{r}),
	\label{eq3}
\end{equation}
$\rho_{k}$ and $E_{k}$ 
%YHK: is 
are obtained with the constraint $eV_b = \mu_{R} - \mu_{L}$. 
Here, $\widehat{h}^{0}_{KS}$, $\Delta v_{Hxc} (\vec{r})$, $\psi_{i} (\vec{r})$, and $\epsilon_{i}$ indicate the ground-state KS Hamiltonian, bias-induced modification of KS potential, KS eigenstates, and KS eigenvalues, respectively. 

\subsection{\label{sec:2-2}Novel features and implementation of MS-DFT}

\begin{figure*}
\includegraphics[width=170mm]{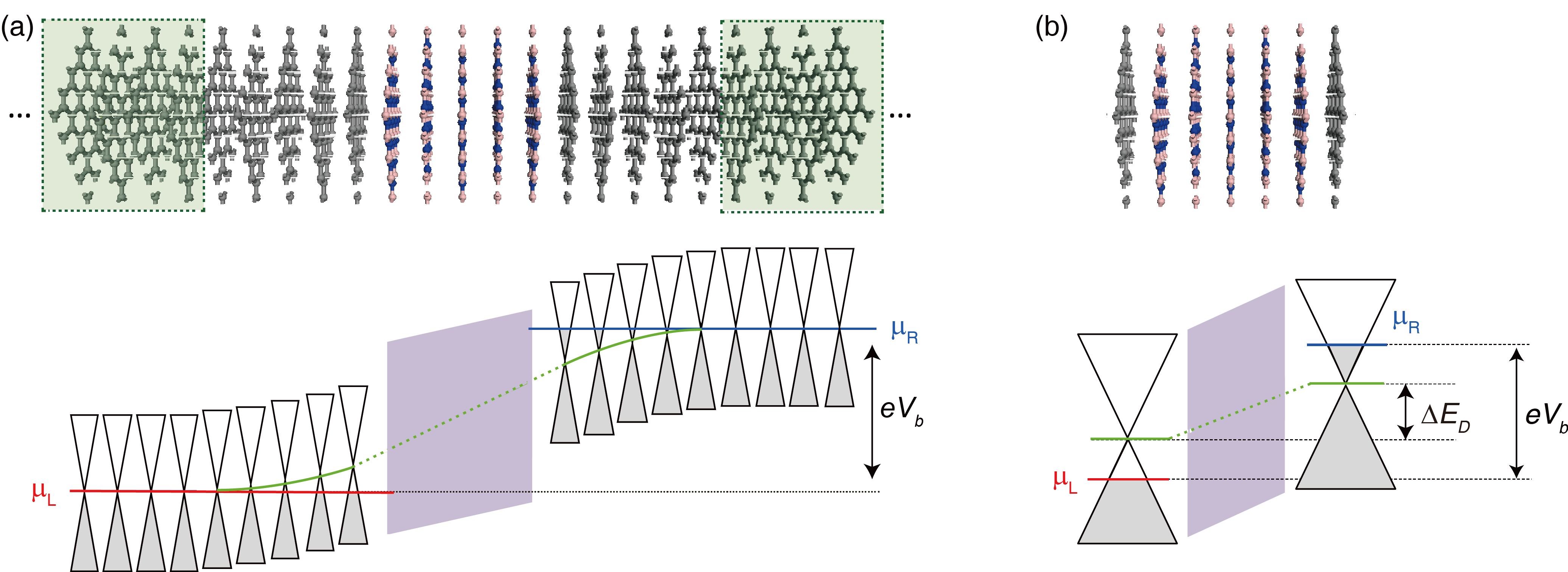}
\caption{Side views of (a) graphite-hBN-graphite and (b) graphene-hBN-graphene vertical heterojunctions (top) and the corresponding schematic energy band diagrams (bottom) at a finite bias voltage $V_b$. Red (blue) line indicates the trace of chemical potential in the left (right) lead and the green line represents the trace of Dirac point across the junction. $\Delta$E\textsubscript{D} indicates the shift of Dirac point of right graphene with respect to that of left graphene.}
\end{figure*}

Recall that within NEGF the matrix elements of the electrode regions are replaced by those of separate bulk calculations in the process of constructing the self-energy $\Sigma_{L(R)}=\tau_{L(R)} g^{L(R)}_{S} \tau_{L(R)}^{\dagger}$, where $\tau_{L(R)}$ is the $L-C$ ($C-R$) coupling matrix and $g^{L(R)}_{S}$ is the $L$ $(R)$ surface Green's function. 
This replacement then directly affects the finite-bias self-consistency cycle for computing the density matrix in NEGF. 
On the other hand, within MS-DFT, the self-consistent cycle for the solution of nonequilibrium KS equations is completed without introducing $\Sigma_{R(L)}$ or the information from separate bulk crystal calculations. 
Instead, only after fully obtaining the nonequilibrium electronic structure, we do so as a post-processing step by invoking the matrix Green’s function formalism \cite{Kim2005,Kim2006} and calculating the transmission function
\begin{equation}
	T(E;V_b)= \\
	Tr\left [ \Gamma_{L} G \Gamma_{R} G^{\dagger} \right ],
	\label{eq4} 
\end{equation}
where $G$ is the retarded Green's function and 
$\Gamma_{L(R)} = i(\Sigma_{L(R)}-\Sigma_{L(R)}^{\dagger})$ 
is the $L$ $(R)$ electrode-induced broadening matrix. 
The current-bias voltage ($I-V_b$) characteristic is then obtained by invoking the Landauer-Buttiker formula \cite{Datta2005,DiVentra2008},
\begin{multline}
	I(V_b)= \\
	\frac{2e}{h}\int ^{\mu_{R}} _{\mu_{L}} T(E;V_b) \left [ f(E - \mu_{R}) - f(E - \mu_{L}) \right ]dE.
	\label{eq5}
\end{multline}

Another important difference between MS-DFT and NEGF is that, unlike NEGF, MS-DFT can naturally treat finite electrodes such as single-layer graphene in vertical vdW heterostructure configuration (\textcolor[rgb]{0,0,1}{Fig. 1\textit{B}}). 
We again emphasize that this is possible because MS-DFT is formally based on the micro-canonical picture. 
In this case, we calculate $T(E;V_b)$ according to \cite{Datta2005,DiVentra2008}
\begin{equation}
	T(E;V_b)=Tr\left[ a_{L} M a_{R} M^{\dagger} \right],
	\label{eq6}
\end{equation}
where $a_{L(R)}$ is the spectral function in the $L$ $(R)$ contact and $M = \tau_{L}^{\dagger} G \tau_{R}$. 
In computing  $a_{L(R)}$, since the semi-infinitely repeated electrode unit cells do not exist anymore for the physically finite electrode case, we replaced the surface Green's function $g_{S}^{L(R)}$ by the region $L$ $(R)$ Green's function $G$ calculated from the junction model with a constant broadening factor $\eta$. The broadening factor,  which originally enters into the construction of $g_{S}$ for the semi-infinite electrode case, physically represents the nature of electrons incoming from (outgoing into) the source (drain) electrode and was set to a value comparable to that used in the semi-infinite electrode case ($\eta \approx 0.025 eV$). 
It should be noted that, while $M$ approximately corresponds to the tunneling matrix in the Bardeen transfer Hamiltonian approach \cite{Bardeen1961}, it now properly accommodates the impact of coupling between the channel and electrodes and their atomistic details \cite{Datta2005}. 

Having physically distinctive metallic electrode/semiconducting channel interfaces, together with a sufficient level of decoupling between $L$ and $R$ states, guarantees the localization (thus assignment) of $\psi_{i}$ into the \textit{L/C/R} regions. 
Since the identification of localized $\psi_{i}$ near $C$ can be generally achieved based on the construction of Wannier functions \cite{Marzari2012}, the spatial assignment of $\psi_{i}$ should be in principle possible irrespective of the choice of basis sets. 
In practice, we implemented MS-DFT within the \texttt{SIESTA} code \cite{Soler2002}, which is based on the linear combination of atomic orbital  formalism and has been extensively employed for the development of DFT-NEGF programs \cite{Taylor2001,Brandbyge2002,Ke2004,Rocha2004}. 
The details of implementing the MS-DFT functionality within an existing DFT code are provided in \textcolor[rgb]{1,0,0}{\textit{SI Appendix}, Section A}. 

\section{\label{sec:3} First-principles descriptions of 2D vertical heterostructure transistors}
 
\subsection{\label{sec:3-1} Negative differential resistance in 2D vertical heterostructure transistors}

\begin{figure*}
\includegraphics[width=170mm]{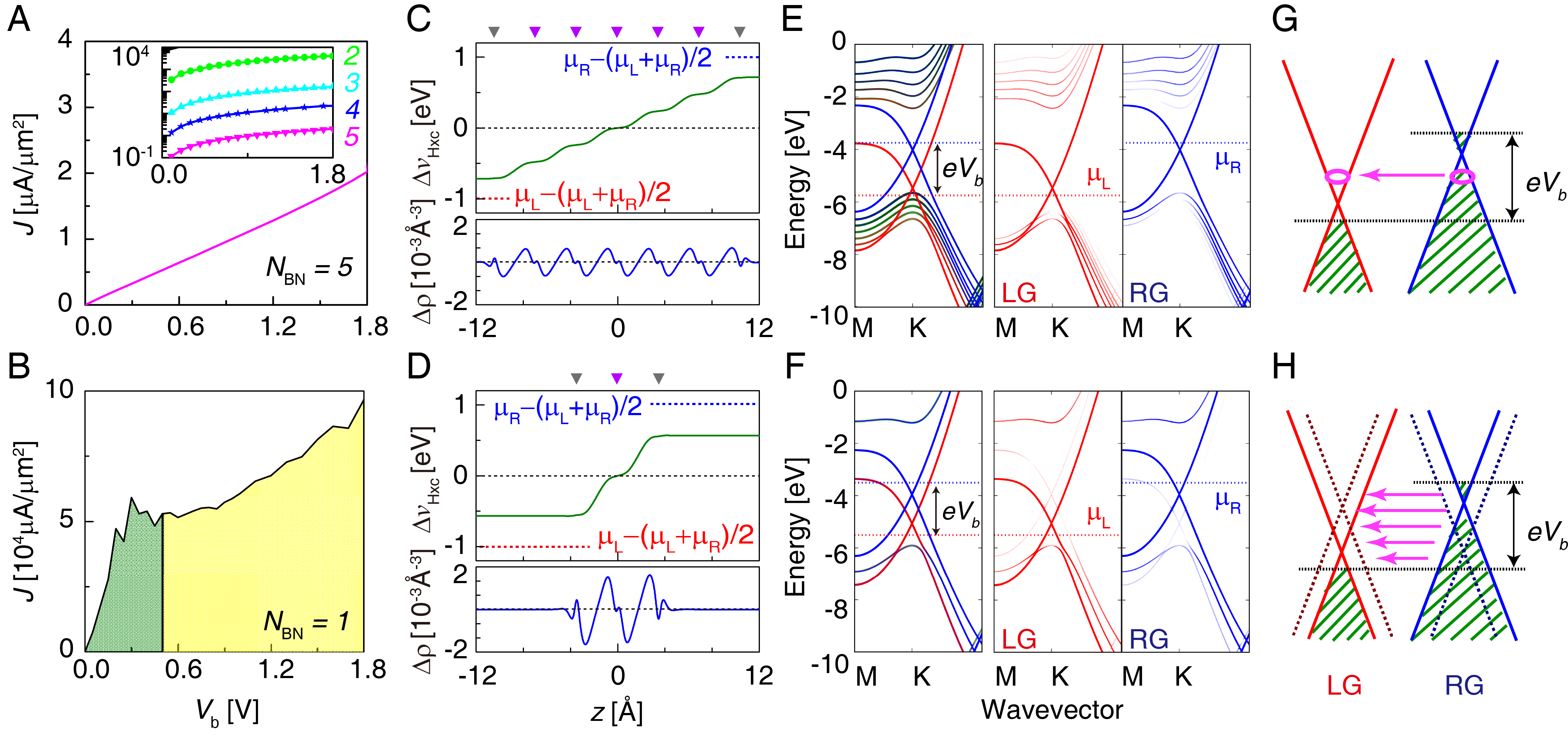}
\caption{The $J-V$ curve of the graphene-hBN-graphene vertical junction for the (a) \textit{N}\textsubscript{BN} = 5 and (b) \textit{N}\textsubscript{BN} = 1 cases. The inset in (a) shows the $J-V$ curves of \textit{N}\textsubscript{BN} from 2 to 5 on a logarithmic scale. 
In (b), the $J-V$ curve shows the low-bias (green shaded area) NDR and high-bias (yellow shaded area) linear current increase. 
$\Delta V\textsubscript{Hxc}$ (upper) and charge density difference (lower) distributions for the (c) \textit{N}\textsubscript{BN} = 5 and (d) \textit{N}\textsubscript{BN} = 1 cases. Gray and purple down triangles indicate the positions of graphene and hBN layers, respectively. Horizontal dotted lines indicate the chemical potentials of LG (red) and RG (blue). 
The projected band structures at $V_b =$ 2.0 V for the (e) \textit{N}\textsubscript{BN} = 5 and (f) \textit{N}\textsubscript{BN} = 1 cases.
The thickness of the band lines quantify the orbital contributions of different 2D layers to the LG (red) and RG (blue) electrodes. 
Schematics of the mechanism of electron transport (magenta arrow) for the (g) \textit{N}\textsubscript{BN} = 5 and (h) \textit{N}\textsubscript{BN} = 1 cases. Red and blue lines indicate the Dirac cones of LG and RG, respectively. Black horizontal dotted lines 	indicate the chemical potentials of LG and RG that define the bias window with the voltage $V_b$.}
\end{figure*}

We apply MS-DFT to study 2D vdW heterostructures composed of a single- or multi-layer hBN sandwiched between graphene electrodes, a prototype configuration for the experimental realization of vertical 2D tunneling transistors and where strong NDR characteristics were observed \cite{Britnell2013}.
We emphasize again that the single-layer graphene electrode case cannot be straightfowardly treated within the conventional NEGF method. 
Instead, for the semi-infinite graphite electrode case, we confirmed that MS-DFT reproduces the finite-bias electronic structure calculated within NEGF (see \textcolor[rgb]{1,0,0}{\textit{SI Appendix}, Section B}). 
Moving on the single-layer graphene electrode cases with the number of hBN layers (\textit{N}\textsubscript{BN}) varying from one to five, we obtain linear $J-V_b$ characteristics for $\textit{N}\textsubscript{BN} \geq 2$ and an exponential current decrease with the increase in the number of hBN layers with a decay rate of 0.718 \AA\textsuperscript{-1} (\textcolor[rgb]{0,0,1}{Fig. 2\textit{A}}). 
On the other hand, for the case of \textit{N}\textsubscript{BN} = 1, we observe a nonlinear current density-bias voltage $(J-V_b)$ behavior (\textcolor[rgb]{0,0,1}{Fig. 2\textit{B}}) characterized by an NDR peak at $V_b = 0.3$ V and a linear current increase at higher voltages.

The notable differences in the $J-V_b$ characteristics of the $\textit{N}\textsubscript{BN} \geq 2$ and $\textit{N}\textsubscript{BN} = 1$ cases can be understood by analyzing their electronic structures with the standard DFT analysis methods, which represents another practical advantage of MS-DFT over NEGF. 
In \textcolor[rgb]{0,0,1}{Fig. 2\textit{C}} and \textcolor[rgb]{0,0,1}{2D}, we show the bias-induced electron density redistribution $\Delta \rho(\vec{r}) = \rho(\vec{r}) - \rho_0(\vec{r}) $ and corresponding $\Delta v_{Hxc}(\vec{r})$ at $V_b = 2.0$ V for the cases of $\textit{N}\textsubscript{BN} = 5$ and $\textit{N}\textsubscript{BN} = 1$, respectively.
For $\textit{N}\textsubscript{BN} = 5$, the offset between left-graphene (LG) and right-graphene (RG) $\Delta v_{Hxc}$ is significantly reduced from that which corresponds to the applied bias (2.0 eV) to 1.45 eV, reflecting the quantum capacitance of graphene or low density of states (DOS) near the graphene Dirac point \cite{Luryi1988, John2004, Fang2007}.
For $\textit{N}\textsubscript{BN} = 1$, the offset of LG and RG $\Delta v_{Hxc}$ is even further reduced to 1.1 eV. 
The corresponding $V_b$-induced $\Delta \rho$ shown together indicates that the amount of interfacial charge transfer in the $\textit{N}\textsubscript{BN} = 1$ case is larger than in the $\textit{N}\textsubscript{BN} = 5$ counterpart, which can be understood by the fact the $\Delta v_{Hxc}$ drop per hBN layer in the former should be larger than that in the latter. 
Projecting the band structures to the LG and RG electrodes as shown in \textcolor[rgb]{0,0,1}{Fig. 2\textit{E}} and \textcolor[rgb]{0,0,1}{2F} also highlights another key distinction between the two cases in that there is no direct hybridization between LG and RG in the $\textit{N}\textsubscript{BN} = 5$ case (only up to second interfacial hBN layer bands are projected onto the graphene bands), the case of $\textit{N}\textsubscript{BN} = 1$ exhibits clear traces of LG Dirac cone on the RG band and vice versa (see also  \textcolor[rgb]{1,0,0}{\textit{SI Appendix}, Section C}). 

\begin{figure*}
\includegraphics[width=170mm]{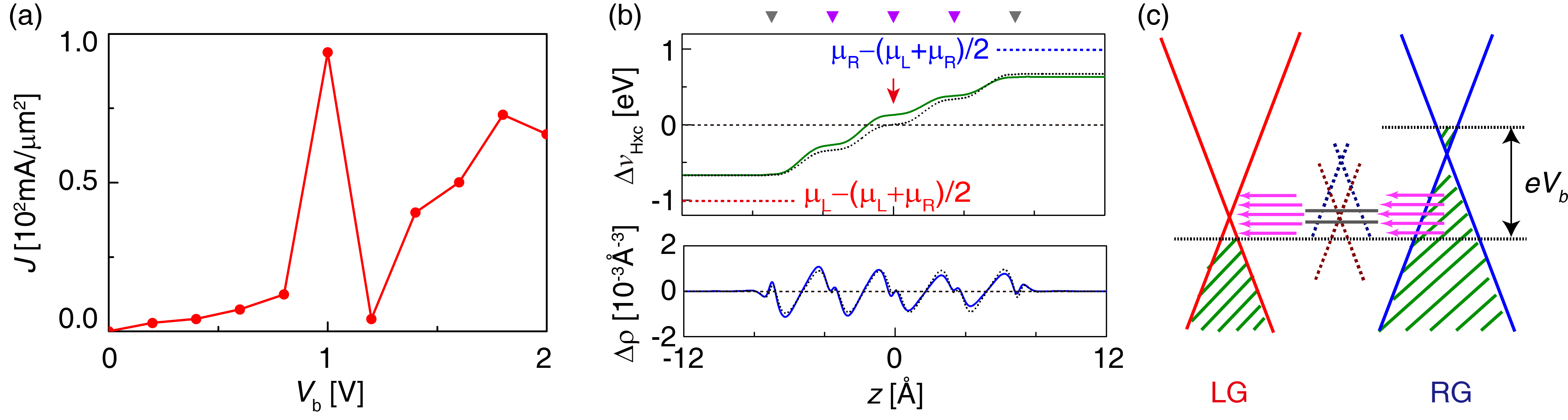}
\caption{(a) Current density-voltage characteristics of graphene−three-layer hBN-single-layer graphene vertical junction with a C\textsubscript{N} defect placed at the central hBN layer. 
(b) $\Delta V\textsubscript{Hxc}$ (upper) and charge density difference (lower) distributions at $V_b = 2.0$ V. 
Gray and purple down triangles indicate the locations of graphene and hBN layers, respectively. 
The red down arrow specifies the location of the defective hBN layer.
Black dotted lines indicate the data obtained for the pristine hBN channel case.
Red and blue horizontal dotted lines in the upper panel indicate the chemical potentials of LG and RG, respectively. 
(c) Schematics of the defect-mediated electron transport near the NDR peak.}
\end{figure*}

Identifying the direct hybridization between the LG and RG states in the $\textit{N}\textsubscript{BN} = 1$ case (\textcolor[rgb]{0,0,1}{Fig. 2\textit{H}}) and no hybridization in the $\textit{N}\textsubscript{BN} \geq 2$ counterparts (\textcolor[rgb]{0,0,1}{Fig. 2\textit{G}}), we now have a basis to understand their distinctively different $J-V_b$ characteristics. 
The $\textit{N}\textsubscript{BN} \geq 2$ cases correspond to the regime where the standard semiclassical treatment, which views the hBN layers as a simple tunnel barrier \cite{Britnell2012_NL, Britnell2012_SC}, is valid. 
Here, $J$ linearly increase with $V_b$ due to the fact that, given the constraint of energy and momentum conservation, tunneling between two shifted graphene Dirac cones at a finite $V_b$ can occur only for a single ring of $\vec{k}$ points and the circumference of these rings linearly increases with $V_b$ (\textcolor[rgb]{0,0,1}{Fig. 2\textit{G}})~\cite{Feenstra2012, Zhao2013}.  
While these semiclassical arguments have been successfully employed to explain several features observed in experiments including the NDR behavior~\cite{Brey2014, Barrera2014, Barrera2015, Greenaway2015, Amorim2016}, atomistic details of hBN and/or graphene are yet to be included. 
On the other hand, the direct hybridization between LG and RG states, 
which could be captured thanks to the first-principles nature of our approach, indicates that the $\textit{N}\textsubscript{BN} = 1$ case is a situation where the details of the potential barrier is important, by opening up transmission eigenchannels that were forbidden within the semiclassical picture (\textcolor[rgb]{0,0,1}{Fig. 2\textit{H}}).

\subsection{\label{sec:3-2} The effect of atomic defects in hBN on NDR}

We next show that the inclusion of atomic defects indeed produces an NDR behavior for few-layer hBN junctions as observed in experiments, demonstrating the capability of our scheme to provide atomistic characterizations of disorder models only phenomenologically introduced in previous semiclassical studies~\cite{Feenstra2012, Zhao2013,Britnell2013}. 
Specifically, we considered a $\textit{N}\textsubscript{BN} = 3$ case and introduced into a central hBN layer a carbon atom substituting a nitrogen atom (C\textsubscript{N}), which is known to be the most preferable defect type for hBN \cite{Azevedo2007}. 
Compared with the pristine $\textit{N}\textsubscript{BN} = 1$ case, the $J-V_b$ characteristics shown in \textcolor[rgb]{0,0,1}{Fig. 3\textit{A}} exhibit not only notably increased currents but also a more pronounced NDR peak at $V_b = 1.0$ V. 
Observing the $\Delta v_{Hxc}$ and $\Delta \rho$ curves at $V = 2.0$ V shown in \textcolor[rgb]{0,0,1}{Fig. 3\textit{B}}, we find that the p-type polarity of C\textsubscript{N} defect states (see \textcolor[rgb]{1,0,0}{\textit{SI Appendix}, Fig. S3}) results in a bigger $\Delta v_{Hxc}$ drop on the drain LG side than the source RG side by 0.18 eV. 
Correspondingly, we observe a bigger $\Delta \rho$ on the former than that on the latter. 
We thus propose that the hybridization between LG and RG states mediated by defect states, as schematically shown in \textcolor[rgb]{0,0,1}{Fig. 3\textit{C}}, is the atomistic origin of NDR in realistic situations.  

\subsection{\label{sec:3-3} The electronic origins of NDR and linear current increase}

\begin{figure*}
\includegraphics[width=170mm]{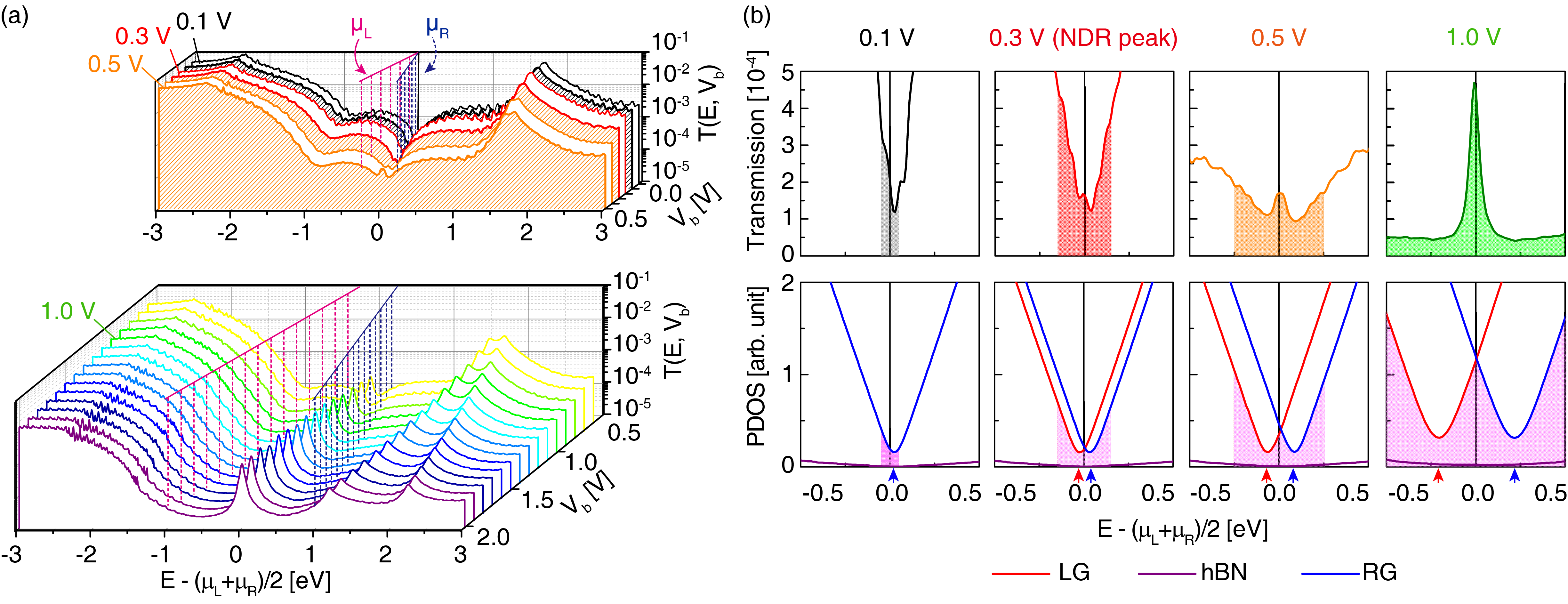}
\caption{(a) The composite of transmission spectra of graphene-hBN-graphene for \textit{N}\textsubscript{BN} = 1 under $V_b$  increasing from 0.0 to 2.0 V at the 0.1 V step. 
Magenta and dark blue lines represent the evolutions of $\mu_{L}$ and $\mu_{R}$ with increasing $V_b$, respectively. 
(b) transmission spectra (upper panels) and PDOS (lower panels) for $V_b =$ 0.1 V (below the NDC peak), 0.3 V (at the NDC peak), 0.5 V (above the NDC peak), and 1.0 V (linear current increase regime). 
In the PDOS plots, red, purple, and blue lines denote the PDOS of LG, hBN, and RG, respectively. 
Shaded areas in $T(E;V_b)$ and PDOS represent the bias windows, and the red and blue up arrows indicate the locations of Dirac points of LG and RG, respectively.}
\end{figure*}

We finally examine the mechanisms of nonlinear $J-V_b$ characteristics of the single-layer hBN channel case in detail (\textcolor[rgb]{0,0,1}{Fig. 2} lower panels). 
In the \textcolor[rgb]{0,0,1}{Fig. 4\textit{A}} upper panel, we show the development of transmission functions $T(E;V_b)$ with increasing $V_b$ for the low-$V_b$ regime where a NDR arises, and in \textcolor[rgb]{0,0,1}{Fig. 4\textit{B}} reproduced the low-bias $T(E;V_b)$ at $V_b = 0.1$ V (below the NDR peak), 0.3 V (NDR peak), and 0.5 V (above the NDR peak), together with the corresponding DOS projected onto the LG and RG electrodes. 
Most notably, due to the above-described LG-RG hybridizations, we observe that the shift of LG and RG Dirac cones is not proportional to $V_b$ and the shape of $T(E;V_b)$ shows a strongly nonlinear behavior. 
Specifically, up to $V_b = 0.3$ V, the shift between LG and RG Dirac points ($\Delta E_D$) minimally increases to 0.06 eV and concurrently $T(E;V_b)$ more or less preserves its zero-bias shape. 
However, as $V_b$ additionally increases from 0.3 V to 0.5 V, $\Delta E_D$ abruptly jumps to 0.14 eV and at the same time $T(E;V_b)$ collapses. 
Quantitatively, the slope of $T(E;V_b)$ at the boundary of bias window (shaded area in \textcolor[rgb]{0,0,1}{Fig. 4\textit{B}}) decreases from 0.005 eV\textsuperscript{-1} at $V_b = 0.1$ V to 0.002 eV\textsuperscript{-1} at $V_b = 0.3$ V and to 0.0006 eV\textsuperscript{-1} at $V_b = 0.5$ V. 

Albeit the NDR peaks observed in experiments could be approximately accommodated within the Bardeen transfer Hamiltonian treatment through the scattering potential or finite coherence models \cite{Feenstra2012, Zhao2013,Britnell2013},
none of the semiclassical approaches could account for the high-bias linear $J$ increase in the measured $J-V_b$ characteristics. 
Being a first-principles approach, MS-DFT indeed successfully produces the linear background current both in the $N\textsubscript{BN} = 1$ (\textcolor[rgb]{0,0,1}{Fig. 2\textit{B}}) and defective $N\textsubscript{BN} = 3$ (\textcolor[rgb]{0,0,1}{Fig. 3\textit{A}}) cases. 
Observing the development of $T(E;V_b)$ at the bias regime $V_b > 0.5$ V in the $N_{\textrm{BN}}=1$ case (\textcolor[rgb]{0,0,1}{Fig. 4\textit{A}} bottom panel), we find that the linear background current originates from the growth in the transmission peak at $(\mu_L+\mu_R)/2$ that has already emerged at $V_b = 0.5$ V (\textcolor[rgb]{0,0,1}{Fig. 4\textit{B}} upper panel). 
This transmission peak was precisely the origin of the linear current increase for the pristine $N_{\textrm{BN}} \geq 2$ cases; namely, our results show that the hybridization between the LG and RG states in the low $V_b$ regime (\textcolor[rgb]{0,0,1}{Fig. 2\textit{H}}) is eventually broken as one increases $V_b$ and the system then switches to a regime where only a single ring of $\vec{k}$ points contributes to the tunneling current (\textcolor[rgb]{0,0,1}{Fig. 2\textit{G}}). 
We thus find that the tunneling mechanisms at low $V_b$ and high $V_b$ compete with each other and the transition between the two will sensitively depend on the atomistic details of hBN barriers, demonstrating the necessity of a first-principle approach to describe the finite-bias quantum transport in 2D vdW heterojunctions.  

\section{Conclusions}
In summary, we established a novel viewpoint that maps a quantum transport process to a multi-space (from drain to source) ``excitation'' within the micro-canonical picture. 
While the viewpoint could be beneficial in a more general context, it particularly allows one to develop a first-principles computational approach based on the constrained-search DFT procedure.
The resulting MS-DFT allows stable numerical calculations of quantum transport phenomena due to its variational nature, and we applied it to 2D vertical  heterostructure transistors. 
Focusing on the single-layer graphene electrode case, which so far could not be treated in a first-principles manner, we showed that the low-$V_b$ NDR behavior arises because of the direct or defect-mediated hybridizations between two graphene electrode states. 
Upon further increasing $V_b$, it was shown that the 2D heterostructure moves into a simple tunneling regime, where one obtains a linear background current. 
The highly nonlinear $J-V_b$ characteristics could not be simultaneously captured in previous semiclassical treatments, which emphasizes the necessity of adopting a first-principles approach to describe 2D vdW heterostructure transistors. 
%The novel atomistic insights emerged here signify that the MS-DFT formalism will make valuable contributions to the development of next-generation devices. 
The novel atomistic insights obtained here demonstrate that the MS-DFT formalism will make valuable contributions to the development of next-generation devices. 

\subparagraph{Methods.}
We adopted the local density approximation~\cite{Ceperley1980}, Troullier-Martins-type norm-conserving pseudopotentials~\cite{Troullier1991}, and numerical atomic orbital basis sets of double-$\zeta$-plus-polarization quality defined with the confinement energy of 100 meV. The 300 Ry of real-space meshgrid cutoff energy and $130 \times 130 \times 1$ Monkhorst-Pack k-point grid were chosen for pristine junction cases where unit cell models are adopted. Single atomic defect C\textsubscript{N} within $5\times5$ supercell hBN, which translates into the defect density of 7.63 × 10\textsuperscript{13}/cm\textsuperscript{2}, and $7 \times 7 \times 1$ Monkhorst-Pack k-point grid were chosen with the inclusion of atomic defect. In calculating $T(E)$, the energy was scanned from $-3.0$ to $+3.0$ eV around $E_{F}$ with 0.01 eV step.

\subparagraph{Author contributions.}
Y.-H.K. oversaw the project and developed the MS-DFT formalism. H.S.K. implemented the method and carried out calculations. Y.-H.K. and H.S.K. analyzed the computational results together and co-wrote the manuscript.

\begin{acknowledgements} This research was supported by the Basic Research Program (NRF-2017R1A2B3009872 and 2017R1A6A3A01011052), the Global Frontier Program (2013M3A6B1078881), and the Nano·Material Technology Development Program (NRF-2016M3A7B4024133) of the National Research Foundation funded by the Ministry of Science, ICT, and Future Planning of Korea. Computational resources were provided by the KISTI Supercomputing Center (KSC-2016-C3-0076 and KSC-2017-C3-0085).
\end{acknowledgements}

\newpage
\bibliographystyle{unsrt}
\bibliographystyle{apsrev4-1}
%\bibliography{MS-DFT_PNAS}

\begin{thebibliography}{39}
	
	\bibitem{Datta2005}
	Datta S (2005) {\em Quantum Transport: Atom to Transistor}.
	\newblock (Cambridge University Press, Cambridge, UK).
	
	\bibitem{DiVentra2008}
	Di~Ventra M (2008) {\em Electrical Transport in Nanoscale Systems}.
	\newblock (Cambridge University Press, Cambridge, UK).
	
	\bibitem{Britnell2012_SC}
	Britnell L, et~al. (2012) Field-effect tunneling transistor based on vertical
	graphene heterostructures.
	\newblock {\em Science} 335(6071):947--950.
	
	\bibitem{Britnell2013}
	Britnell L, et~al. (2013) Resonant tunnelling and negative differential
	conductance in graphene transistors.
	\newblock {\em Nat Commun} 4:1794.
	
	\bibitem{Jena2013}
	Jena D (2013) Tunneling transistors based on graphene and 2-d crystals.
	\newblock {\em Proc IEEE} 101(7):1585--1602.
	
	\bibitem{Li2016}
	Li MY, Chen CH, Shi Y, Li LJ (2016) Heterostructures based on two-dimensional
	layered materials and their potential applications.
	\newblock {\em Mater Today} 19(6):322--335.
	
	\bibitem{Iannaccone2018}
	Iannaccone G, Bonaccorso F, Colombo L, Fiori G (2018) Quantum engineering of
	transistors based on 2d materials heterostructures.
	\newblock {\em Nat Nano} 13(3):183.
	
	\bibitem{Mak2010}
	Mak KF, Sfeir MY, Misewich JA, Heinz TF (2010) {The evolution of electronic
		structure in few-layer graphene revealed by optical spectroscopy}.
	\newblock {\em Proc Natl Acad Sci U S A} 107(34):14999--15004.
	
	\bibitem{Kang2016}
	Kang S, et~al. (2016) Effects of electrode layer band structure on the
	performance of multilayer graphene-hBN-graphene interlayer tunnel field
	effect transistors.
	\newblock {\em Nano Lett} 16(8):4975--4981.
	
	\bibitem{Bardeen1961}
	Bardeen J (1961) Tunnelling from a many-particle point of view.
	\newblock {\em Phys Rev Lett} 6(2):57--59.
	
	\bibitem{Tersoff1985}
	Tersoff J, Hamann DR (1985) Theory of the scanning tunneling microscope.
	\newblock {\em Phys Rev B} 31(2):805--813.
	
	\bibitem{Feenstra2012}
	Feenstra RM, Jena D, Gu G (2012) Single-particle tunneling in doped
	graphene-insulator-graphene junctions.
	\newblock {\em J Appl Phys} 111(4):043711.
	
	\bibitem{Zhao2013}
	Zhao P, Feenstra RM, Gu G, Jena D (2013) Symfet: A proposed symmetric graphene
	tunneling field-effect transistor.
	\newblock {\em IEEE Trans Electron Dev} 60(3):951--957.
	
	\bibitem{Campbell2015}
	Campbell PM, Tarasov A, Joiner CA, Ready WJ, Vogel EM (2015) Enhanced resonant
	tunneling in symmetric 2d semiconductor vertical heterostructure transistors.
	\newblock {\em ACS Nano} 9(5):5000--5008.
	
	\bibitem{DiVentra2004}
	Di~Ventra M, Todorov TN (2004) Transport in nanoscale systems: the
	microcanonical versus grand-canonical picture.
	\newblock {\em J Phys Condens Matter} 16(45):8025.
	
	\bibitem{Bushong2005}
	Bushong N, Sai N, Di~Ventra M (2005) Approach to steady-state transport in
	nanoscale conductors.
	\newblock {\em Nano Lett} 5(12):2569--2572.
	
	\bibitem{Levy1999}
	Levy M, Nagy \'A (1999) Variational density-functional theory for an individual
	excited state.
	\newblock {\em Phys Rev Lett} 83(21):4361--4364.
	
	\bibitem{Ayers2009}
	Ayers PW, Levy M (2009) Time-independent (static) density-functional theories
	for pure excited states: Extensions and unification.
	\newblock {\em Phys Rev A} 80.
	
	\bibitem{Gorling1999}
	G\''orling A (1999) Density-functional theory beyond the hohenberg-kohn theorem.
	\newblock {\em Phys Rev A} 59:3359--3374.
	
	\bibitem{Kim2005}
	Kim YH, Jang SS, Jang YH, Goddard WA (2005) First-principles study of the
	switching mechanism of [2]catenane molecular electronic devices.
	\newblock {\em Phys Rev Lett} 94(15):156801.
	
	\bibitem{Kim2006}
	Kim YH, Tahir-Kheli J, Schultz PA, Goddard WA (2006) First-principles approach
	to the charge-transport characteristics of monolayer molecular-electronics
	devices: Application to hexanedithiolate devices.
	\newblock {\em Phys Rev B} 73(23):235419.
	
	\bibitem{Marzari2012}
	Marzari N, Mostofi AA, Yates JR, Souza I, Vanderbilt D (2012) Maximally
	localized wannier functions: Theory and applications.
	\newblock {\em Rev Mod Phys} 84(4):1419.
	
	\bibitem{Soler2002}
	Soler JM, et~al. (2002) The siesta method for ab initio order-n materials
	simulation.
	\newblock {\em J Phys Condens Matter} 14(11):2745.
	
	\bibitem{Taylor2001}
	Taylor J, Guo H, Wang J (2001) Ab initio modeling of quantum transport
	properties of molecular electronic devices.
	\newblock {\em Phys Rev B} 63(24):1--13.
	
	\bibitem{Brandbyge2002}
	Brandbyge M, Mozos JL, Ordej\'on P, Taylor J, Stokbro K (2002)
	Density-functional method for nonequilibrium electron transport.
	\newblock {\em Phys Rev B} 65(16):165401.
	
	\bibitem{Ke2004}
	Ke SH, Baranger HU, Yang W (2004) Electron transport through molecules:
	Self-consistent and non-self-consistent approaches.
	\newblock {\em Phys Rev B} 70(8):085410.
	
	\bibitem{Rocha2004}
	Rocha A, Sanvito S (2004) Asymmetric I-V characteristics and magnetoresistance
	in magnetic point contacts.
	\newblock {\em Phys Rev B} 70(9):094406.
	
	\bibitem{Luryi1988}
	Luryi S (1988) Quantum capacitance devices.
	\newblock {\em Appl Phys Lett} 52(6):501--503.
	
	\bibitem{John2004}
	John DL, Castro LC, Pulfrey DL (2004) Quantum capacitance in nanoscale device
	modeling.
	\newblock {\em J Appl Phys} 96(9):5180--5184.
	
	\bibitem{Fang2007}
	Fang T, Konar A, Xing H, Jena D (2007) Carrier statistics and quantum
	capacitance of graphene sheets and ribbons.
	\newblock {\em Appl Phys Lett} 91(9):092109.
	
	\bibitem{Britnell2012_NL}
	Britnell L, et~al. (2012) Electron tunneling through ultrathin boron nitride
	crystalline barriers.
	\newblock {\em Nano Lett} 12(3):1707--1710.
	
	\bibitem{Brey2014}
	Brey L (2014) Coherent tunneling and negative differential conductivity in a
	graphene/$h$-bn/graphene heterostructure.
	\newblock {\em Phys Rev Applied} 2(1):014003.
	
	\bibitem{Barrera2014}
	de la Barrera SC, Gao Q, Feenstra RM (2014) Theory of
	graphene-insulator-graphene tunnel junctions.
	\newblock {\em J Vac Sci Technol B} 32(4):04E101.
	
	\bibitem{Barrera2015}
	de la Barrera SC, Feenstra RM (2015) Theory of resonant tunneling in
	bilayer-graphene/hexagonal-boron-nitride heterostructures.
	\newblock {\em Appl Phys Lett} 106(9):093115.
	
	\bibitem{Greenaway2015}
	Greenaway M, et~al. (2015) Resonant tunnelling between the chiral landau states
	of twisted graphene lattices.
	\newblock {\em Nat Phys} 11(12):1057.
	
	\bibitem{Amorim2016}
	Amorim B, Ribeiro RM, Peres NMR (2016) Multiple negative differential
	conductance regions and inelastic phonon assisted tunneling in
	graphene-bn-graphene structures.
	\newblock {\em Phys Rev B} 93(23):235403.
	
	\bibitem{Azevedo2007}
	Azevedo S, Kaschny JR, de~Castilho CMC, de~Brito~Mota F (2007) A theoretical
	investigation of defects in a boron nitride monolayer.
	\newblock {\em Nanotechnology} 18(49):495707.
	
	\bibitem{Ceperley1980}
	Ceperley DM, Alder BJ (1980) Ground state of the electron gas by a stochastic
	method.
	\newblock {\em Phys Rev Lett} 45(7):566--569.
	
	\bibitem{Troullier1991}
	Troullier N, Martins JL (1991) Efficient pseudopotentials for plane-wave
	calculations.
	\newblock {\em Phys Rev B} 43(3):1993--2006.
	
\end{thebibliography}

\end{document}